# Discursive Landscapes and Unsupervised Topic Modeling in IR: A Validation of Text-As-Data Approaches through a New Corpus of UN Security Council Speeches on Afghanistan

Mirco Schönfeld (Bavarian School of Public Policy at TU Munich), Steffen Eckhard (University of Konstanz), Ronny Patz (LMU Munich), Hilde van Meegdenburg (Leiden University)

**Abstract**

The recent turn towards quantitative text-as-data approaches in IR brought new ways to study the discursive landscape of world politics. Here seen as complementary to qualitative approaches, quantitative assessments have the advantage of being able to order and make comprehensible vast amounts of text. However, the validity of unsupervised methods applied to the types of text available in large quantities needs to be established before they can speak to other studies relying on text and discourse as data. In this paper, we introduce a new text corpus of United Nations Security Council (UNSC) speeches on Afghanistan between 2001 and 2017; we study this corpus through unsupervised topic modeling (LDA) with the central aim to validate the topic categories that the LDA identifies; and we discuss the added value, and complementarity, of quantitative text-as-data approaches. We set-up two tests using mixed-method approaches. Firstly, we evaluate the identified topics by assessing whether they conform with previous qualitative work on the development of the situation in Afghanistan. Secondly, we use network analysis to study the underlying social structures of what we will call 'speaker-topic relations' to see whether they correspondent to know divisions and coalitions in the UNSC. In both cases we find that the unsupervised LDA indeed provides valid and valuable outputs. In addition, the mixed-method approaches themselves reveal interesting patterns deserving future qualitative research. Amongst these are the coalition and dynamics around the 'women and human rights' topic as part of the UNSC debates on Afghanistan.

**Acknowledgements**

*Early stages of this work have greatly profited from the input of Alena Moiseeva (Center for Information and Language Processing (CIS), LMU Munich) without whom we might not have embarked on this endeavor. We thank Patrick A. Mello (Bavarian School of Public Policy) for sharing his Afghanistan expertise and colleagues at the LMU Munich International Relations Colloquium and colleagues at the ECPR General Conference Section: 'Political Networks', Panel: 'From Texts to Networks: Semantic, Socio-Semantic and Discourse Networks' for constructive comments and suggestions on earlier versions of this paper. Thanks also go to Niheer Dasandi (University of Birmingham) for advice on how to focus this text further.*

**Dataset and interactive visualization**

*The dataset of UNSC speeches on the situation in Afghanistan and additional metadata required for replication is available online (*https://doi.org/10.7910/DVN/OM9RG8*). An interactive visualization of our findings is also online (*https://mirco.shinyapps.io/unscafg/*).*

# 1. Introduction

In this paper we introduce unsupervised quantitative text analysis and speaker-topic networks as tools to study the discursive landscape of debates in the United Nations (UN), in particular in the UN Security Council (UNSC). We build on the growing literature on quantitative text analysis in international relations (IR) and advance ongoing debates by placing the focus on validating the methods used through a mixed-method approach. Therewith, this paper presents new applications of text-as-data analyses and innovative ways to analyze the large corpora of speech and text produced by international organizations. We introduce a new text corpus encompassing all public UNSC debates on Afghanistan during the period 2001-17[1], consisting of 2347 individual speeches by officials representing states, UN organizations and non-state actors, and we analyze this corpus using Latent Dirichlet Allocation (LDA) and network analysis. LDA is a text analysis tool that automatically assigns topics to texts (Blei et al. 2003), allowing for an analysis without prior knowledge of, or assumptions about, the content of the texts. Based on the assignment of topics to text, in our case individual speeches at UNSC debates, we can trace the evolution of topics over time, including their relative relevance, and analyze the structure of the speaker-topic network that underlays it. Such a speaker-topic network is not only tool for validation, it also allows us to study social relations and resulting networks through text providing new ways to visualize the landscape of an international discourse over an extend period of time.

Studying the UN debates on Afghanistan and analyzing the structure of the conflict as represented by such debates has been, until recently, a domain for qualitative research – in particular qualitative discourse analysis (e.g. Donahue and Prosser 1997; Patil 2009; Puechguirbal 2010; McDonald 2013; Quinton-Brown 2013). However, recent advances in quantitative text analysis in IR, especially since King and Lowe (2003), have led to a variety of contributions using "text-as-data" approaches (Grimmer and Stewart 2013; Wilkerson and Casas 2017). These contributions show that quantitative text analysis can yield significant new insights for the study of international relations. For instance, the recent creation and analysis of new text corpora (a notable example is the UN General Debate corpus created by Baturo, Dasandi and Mikhaylov (2017)) already provided novel insights regarding the UN system and international politics more broadly (Gurciullo and Mikhaylov 2017a, 2017b; Pomeroy, Dasandi, Mikhaylov 2018).[2] One of the main advantages of text-as-data approaches thereby is their ability to present and structure large amounts of text without making any *a priori* assumptions or selections. This distinguishes quantitative text-as-data approaches from qualitative approaches were cases (topics or countries) are selected because of their expected relevance or their relative power position. Quantitative approaches allow researchers to identify relevant speakers and topics beyond those that are most frequently studied (i.e. the Permanent Five (P5) of the UNSC), often with a strong US-focus (Puchala 2005, 574).

---

[1] Version 1 of 'The UN Security Council debates on Afghanistan' corpus (Schönfeld, Eckhard, Patz, van Meegdenburg, 2018) and related files are available at https://doi.org/10.7910/DVN/OM9RG8.

However, considering text-as-data approaches are relatively new to IR it is important to ensure these methods are sufficiently validated. Whilst previous studies have shown findings to be reliable and broad in scope, in this paper we focus on cross-validating LDA-based topics with qualitative expert assessments and with an analysis of resulting speaker-topic networks. By these means we can assess the degree to which the outcome of the quantitative analysis corroborates, and is corroborated by, existing knowledge about the conflict in Afghanistan and existing knowledge about inter-state relations.

The contribution this article makes is threefold. Firstly, it advances quantitative text analysis in IR by presenting and analyzing a new text corpus and by illustrating how one can distill the underlying 'discursive landscape' from such a corpus. Secondly, taking the UNSC debates on Afghanistan as a case study, it demonstrates that automated text analysis tools can produce reliable, valid and novel results. The reliability and validity will be demonstrated by corroborating the outcome of an LDA-topic modeling with qualitative descriptions of the conflict and by analyzing the country coalitions (networks) that table certain topics in the UNSC. To illustrate the novelty of the results, we will show how the results are distinct from, yet complement, both more constructivist, qualitative discourse analyses (see below) and more positivist approaches such as Lagassé's and Mello's (2018) analysis of media coverage of

---

[2] For a methodological discussion see: Kutter (2018).

Afghanistan and Afghan policy. Thirdly, beyond demonstrating the use of quantitative discursive approaches, we discuss the theoretical implications and overall methodological potential of being able to analyze large text corpora.

The paper is structured as follows. We first address the added value and complementarity of quantitative text analysis to traditional discursive approaches. This section explains that although we take a different methodological approach to studying discourse we share the main assumptions and understanding of the relevance of discourse for the study of IR. Most importantly, what we aim to argue here is that the type of quantitative approaches introduced in this paper should not be seen as challenging but as complimentary to qualitative discursive approaches. Thereafter, in section 3, we introduce the new text-corpus in detail and elaborate on the steps we took to prepare this corpus for analysis. In section 4 we explain in greater detail the LDA-based topic assignment and elaborate on the topic-categories that derive from this method. The two subsequent sections, sections 5 and 6, then focus on the external validation of the method in two steps. In section 5 we draw on qualitative studies of the conflict in Afghanistan to show that LDA-based topic modeling can indeed capture the overall development of the conflict and intervention over time, including major events in the period studied. In section 6 we combine the LDA-topics with a topic-speaker network analysis to show how particular, and known, coalitions of states focus on particular topics. To be precise, we will analyze the networks surrounding two topics identified through the unsupervised method: drugs and women/human rights. We close with reflections on the method and its (dis)advantages.

2. **From discourse analysis to quantitative text analysis: mapping the discursive landscape through text-as-data approaches**

Studying language and discourse has a long tradition in IR. Qualitative discourse analysis recognizes the performative and social functions of discourse, whereby in particular the processes through which discourses construct and reproduce meanings, identities and power relations grew (cf. Milliken, 1999). In international security studies, discourse analysis made possible the study and apprehension of securitization practices (Buzan, Weaver and de Wilde 1998; Huysmans 2006); of narratives as enabling, justifying and legitimating particular courses of action, including military interventions (Spencer 2018); and the reproduction of particular gendered and racial social structures shaping state and military practices (Shepherd 2008;

Ayotte and Husain 2005; Krook and True 2010). Although different strands of research may emphasize different aspects of discursive processes and practices there is a general agreement within constructivist approaches that discourses are both an expression of subjective positions and actor identities, simultaneously shaped by, shaping and reproducing social relations, and constitutive of spaces of action and possibility (Jackson 2007). The type of quantitative analysis conducted in this paper is not meant to replace these approaches to discourse analysis. Instead, it complements and may strengthen the tools we have for working with text, making it easier to identify broader trends, both over time and across fora; allowing for the selection of interesting cases and narratives for in-depth analysis; or to cross-validate findings from earlier research. Moreover, whilst quantitative text-as-data approaches themselves are ontologically agnostic – the main strength of these approaches lays in their ability to identify patterns and foci, not in evaluating causal claims – we largely share two major assumption underlying most qualitative approaches to discourse.

Firstly, we look at discourses as reflections of state subjective positions. We assume state foreign policy foci are not just predicated on state survival or a function of economically defined self-interests but shaped by historically grown identities as 'situated subjectivities': 'one's sense of who one is, of one's social location, and how (given the first two) one is prepared to act' (Brubaker and Cooper 2000:17). With respect to collective actors, such as states, shared understandings of a state's identity and social position are neither fixed over time nor, in most cases, domestically uncontested but they are found to inform how state interests are defined and what foreign policy is pursued (Guzzini 2012; Bevir and Daddow 2015; Bucher and Jasper 2017; van Meegdenburg 2018). Hence, besides domestic concerns and a state's perception of its geopolitical position, we assume state foreign policy is informed by the dominant ideational structures in a country and the norms and values it adheres to. Therewith, apart from non-discursive behavioral outcomes (i.e. voting records, participation in interventions), discourse and speech acts form a central entrance to studying and understanding state positions: they tell us what a state stands for and what is perceived as important.

Secondly, we study the UNSC debates because these debates set the boundaries of the politically feasible. Although itself a highly exclusionary political body (Morris 2000), the UNSC debates preceding the resolutions relating to the mandate of the UN Assistance Mission in Afghanistan (UNAMA), and other meetings and debates that specifically address the situation in Afghanistan, ultimately frame what is perceived as important and delimit the space of collective action. This explains the relevance of UNSC membership for states, and the politics and contestation surrounding the P5 and the elections of the rotating members (Bourantonis 2005; Stuenkel 2010; Maseng and Lekaba 2014), but it also underscores the relevance of conducting holistic (longitudinal) analysis of the UNSC debates regarding Afghanistan. The latter because, beyond the final wording of the resolutions and the voting patterns of states, the debates preceding the resolutions give a more nuanced view of the salience of different topics and the different foci states bring in.

Following these assumptions is neither necessary nor can they be 'proven' by the analysis. However, as qualitative and quantitative approaches are often discussed as in opposition with one another – and since research tends to do either one or the other – we find it important to relay our conviction that, in this case, qualitative and quantitative approaches are *complementary*. Each approach has particular strengths that may aid, and be aided by, analysis at the other level. More specifically, whilst qualitative discourse analyses frequently focus on how particular narratives and interpretations render particular courses of action (im)possible (Spencer 2018, 31-2), a quantitative analysis can better describe the overall discursive landscape focusing on the salience of particular topics with the same aim.

To date, discursive approaches to the intervention in Afghanistan focused on the construction of consent for the intervention in a single country (Boucher, 2009; Jakobsen and Ringsmose 2015; Kriner and Wilson 2016) or on the discourse and position of single, influential people (Kerton-Johnson 2008; Reyes 2011; McCrisken 2012). Alternatively, studies tended to focused on single topics. Examples are discourse analyses of the particular understandings of development and state-building (Naylor 2011), of Afghan women and the discursive construction of their social position (Ayotte and Husain 2005) or of the Taliban insurgency (Johnson 2007). Such focused, qualitative analysis are and remain important – not in the least since states, when talking about common issues, may refer to common norms and values but actually mean and relate to very different 'things' (Pouliot and Thérien, 2017). However, offering the advantage of in-depth analysis, these approaches necessarily forego embedding these specific discourses in the overall, international discourse on Afghanistan. The quantitative analysis presented in this paper takes a different approach. It sets out to map the overall 'discursive landscape' as it is constructed by multiple actors in relation to all topics deemed relevant by them. It therewith compliments qualitative approaches and allows us to situate the different foci relative to one another. For instance, rather than looking at the particular construction of 'womanhood' in Afghanistan, a quantitative analysis can reveal how prominently the position of women was discussed in the UNSC and which actors were mostly concerned with this question at which point in the intervention.

Overall, the main advantage of computer-based content analysis is that it enables the study of speech and discourse beyond the point where qualitative approaches reach their limits. "[C]omputational international relations" (Akin Unver 2018) is developing tools to map the debates and contributions of numerous actors, over longer periods of time and across different fora. Using some of these tools we can map the discursive landscape in relation to the intervention in Afghanistan as created by the contributions of all UNSC members, including the P5, the different rotating members, guests and UN officials since the beginning of the intervention. This landscape not only gives us an impression of what was going on 'on the ground' but also of the different foreign policy foci of the members – it tells us what was deemed important, when and by whom. Different from qualitative studies into country foreign policy identities, this allows us to identify country particular foci as well as the overall emphasis on particular topics within the broader discourse. In the end, although this paper is descriptive in nature, mapping and plotting foreign policy foci and networks through discourse, we do so specifically because of what these debates and foci enable: the UNSC shapes what type of action, in relation to which identified problems, becomes possible and feasible.

*Expectations and means of validation* – In this paper we focus on the UN and the UNSC. Yet, the methods introduced should be broadly applicable and advance the study of IR through text corpora in general. Therefore, besides demonstrating the usefulness and complementarity of quantitative approaches, we focus on validating their findings.

Although there is a large methodological literature addressing the validity of language processing techniques, the validity of topic modelling techniques remains problematic. This is particularly related to the interpretation of the topic categories produced by probabilistic topic models (Chang et al. 2009) – i.e. what do they signify and what should they signify? To overcome this and to allow for validation of the automated topic analysis we will assess the outcome 'topics' by two means: 1) an assessment of the topics and their relative relevance over time in relation to qualitative accounts of the conflict; and 2) an assessment of the topic-speaker networks that underlie the debates. For each of these means of validation we can formulate an expectation:

> ***Hypothesis 1****: Shifts in dominance of specific Afghanistan-related topics in UNSC speeches should correspond to key events and shifts from key periods identified by previous research on Afghanistan.*

This hypothesis would be (partially or fully) falsified if the automatic topic assignment of speeches and the changes of dominant topics over time does not correspond to key events and periods (e.g. elections or power transitions) in the conflict. To verify this, we look at topics identified in earlier qualitative work on the conflict and UN mission in Afghanistan.

> ***Hypothesis 2****: The structure of speaker-topic relations represents the traditional or otherwise expected divides in the UNSC and the security related topics these actors deem relevant.*

This hypothesis would be (partially or fully) falsified if the speaker-topic structure (further explained below) resulting from the automated text analysis does not show any variation in the positions of UNSC members, or would show that they ascribe equal relevance to all topics identified, or would not cluster at least broadly around known geopolitical divisions.

In relation to both expectations the underlying logic of validation is the following. Based on previous knowledge about the conflict in Afghanistan and based on previous knowledge about state relations and inter-state networks and coalitions we assess whether we can observe expected patterns in the automated 'outputs' of the quantitative analysis. In the first step the automated output is the topics assigned to the speeches. In the second step the output is the network analysis based on these topics. In the following four sections, we first present the new text corpus comprising all UNSC speeches on Afghanistan between 2001 and 2017 (section 3) and discuss the method by which we applied automated text analysis (section 4). We then present the findings regarding the evolution of topics and their relevance over time and discuss, in line with H1, their validity given the state of the art in relation to the conflict and intervention in Afghanistan (section 5). We subsequently present our findings on the structure of the speaker-topic relations and, following H2, discuss the extent to which they align with expected patterns (section 6). The concluding section, section 7, then summarizes our findings on the validity of the tool – the question whether an LDA-based topic modeling can indeed capture the major topics in relation to such a complex conflict environment as Afghanistan – and addresses the comparative (dis)advantages of the method.

### 3. Presenting the UN Security Council Debates on Afghanistan corpus: automatic speech extraction with human checking and additional coding

To compile the UNSC speech corpus we downloaded all publicly available transcripts of meetings of the UNSC debating the situation and intervention in Afghanistan between 2001-2017 from the website of the UN.[3] In order to compile a corpus of single speeches from these UNSC meeting protocols, we extracted speeches by individual countries, UN organizations and third-party actors. Given the focus of this paper, we selected only those protocols labeled with "Situation in Afghanistan" or "Afghanistan". Speeches were split up automatically from the respective protocols. The initial speech extraction algorithm was validated by checking all individual speech files of the corpus to see whether they contained more than one speech. If they contained more than one speech the algorithm was adapted, which resulted in improved algorithmic speech recognition. This reiterated procedure provided us with a total of 2347 individual speeches (see *Table 1*), consisting of between 55 and 201 speeches per year (see *Figure 1*).

*Table 1. Descriptive statistics for the overall corpus*

| **Description** | **Characteristic** |
|---|---|
| Timespan of protocols | June 5, 2001 – December 21, 2017 |
| Number of Protocols | 130 |
| Number of Countries / Affiliations | 103 |
| Number of Speeches | 2,347 |

---

[3] http://www.un.org/en/sc/meetings/ (last accessed: October 2018)

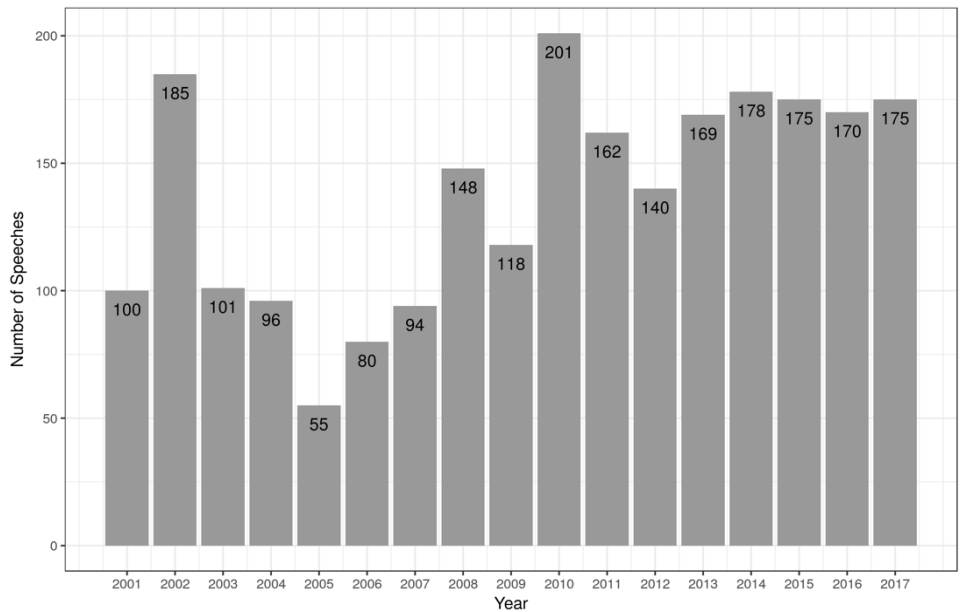

*Figure 1. Distribution of speeches per year*

Each speech was then labelled with speaker affiliation information (e.g. China, UN etc.) based on the list of attendees provided at the beginning of each meeting protocol. Where speaker names could not be affiliated with a country or an UN-institution, the final list of all speakers was checked for their country or organizational affiliation at time of speaking. For 31 speakers who were not listed among the attendees we had to enter their affiliation manually. For example, one of these speakers was Mr. Nicholas Haysom who spoke in his capacity as Special Representative of the Secretary-General for Afghanistan and Head of UNAMA while his speech was indicated by "Mr. Haysom" only. In case of Mr. Haysom, we assigned "UN" as his affiliation. The final dataset includes 103 distinct speaker affiliations. The dataset analyzed in this paper excludes all interventions made by "The President" of the UNSC, because distinguishing the president speaking in her or his presidential role or national capacity was not always possible.

This means that, whilst most of the extraction of single speeches can be done by automated algorithmic speech extraction, human intervention remains necessary to ensure accurate speaker-affiliation labelling of all speakers. However, this is due to the fact not all affiliations were clearly stated in the protocols. *Figure 2* shows the total number of speeches per year for the ten most 'active' overall speakers.

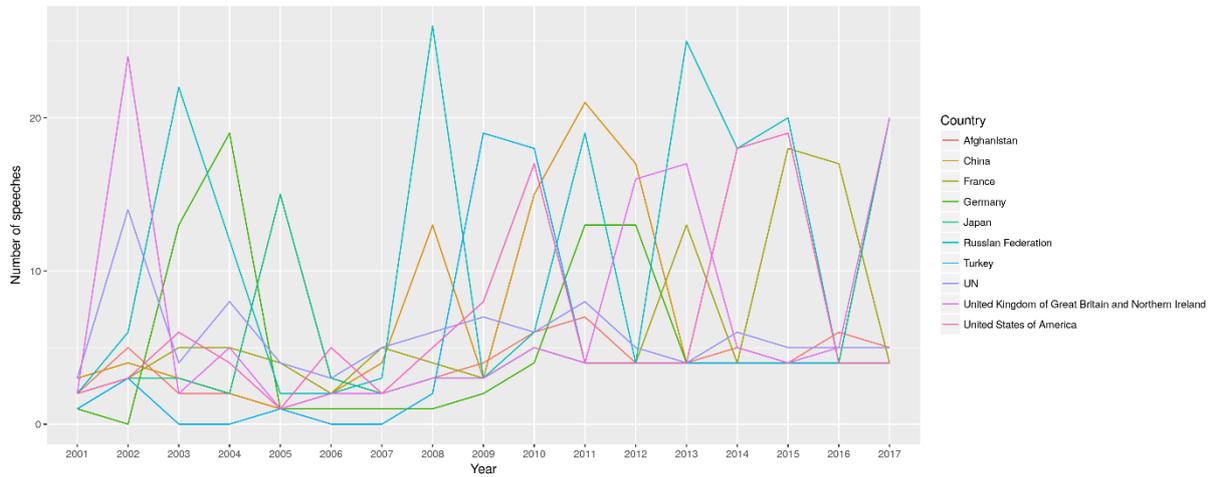

*Figure 2. Number of speeches per year for the ten countries with the highest number of speeches in the corpus*

The entire corpus, consisting of the above-described set of distinct speeches together with the associated metadata (speakers' names and affiliations), is available online (Schönfeld, Eckhard, Patz, van Meegdenburg, 2018). In the following, we first discuss how we applied LDA-based topic modeling. Thereafter, we focus on validating the topic-categories we derive from this method.

4. **Presenting the discursive landscape: LDA-based topic assignment and speaker-topic networks**

Considering each of the 2,347 speeches has two anchors – year and speaker/country affiliation – we were able to conduct a topic modeling emphasizing changes over time as well as laying bare the particular emphasis different speakers position in the speaker-topic network. In a first step, we applied common means of pre-processing such as tokenization of text, followed by lemmatization of tokens, creation of a vocabulary and pruning of vocabulary. Pruning includes removal of typical stop-words as well as seldom words. We decided to remove words from the vocabulary that occur less than three times.

Based on this preprocessed text, we used an algorithm for LDA proposed by Blei et al. (2003). Given that this algorithm does not select a definite number of topics, we subsequently applied a metric described by Deveaud et al. (2014) that quantifies the divergence between all pairs of LDA topics produced in one execution of the procedure to choose the optimal number of topics to describe the corpus. By iteratively executing LDA, varying the number of topics $k$ for each iteration and calculating the before-mentioned metric, one can find the $k$ that results in a topic model with maximal expressiveness in terms of information divergence between all pairs of topics. We applied this iterative approach to the dataset varying $k$ to take values between 2 and 25. The corresponding results reported by the Deveaud-metric are depicted in *Figure 3*. We validated the possible topic assignments first individually and then jointly in two 2-3h sessions to find that a higher number of topics resulted in too many topic artefacts that could not be identified as distinct topics, suggesting that choosing the first local peak in the Deveaud-metric appeared to produce a consistently valid list of topics. This means that, for our analysis, we worked with $k = 10$.

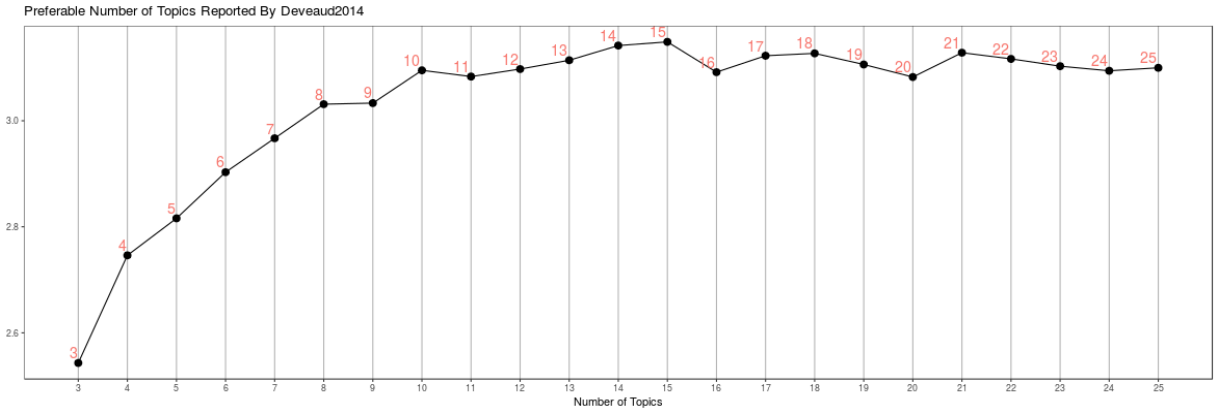

*Figure 3 Evaluation of topic models for different numbers of topics based on Deveaud metric.*

Once a number of topics is selected – 10 topics in our case – one can map the temporal changes in the relative relevance of the topics for the full time period of the corpus (see *Figure 4*). Considering individual speeches in the corpus are subsumed under a particular topic-heading depending on the extent to which they address that particular topic, this landscape depicts both the relative dominance of a particular topic in a given year and the evolution of the debate over time. From *Figure 4* we can see that this the discursive landscape of the situation and intervention in Afghanistan shows significant variation in dominant topics over time.

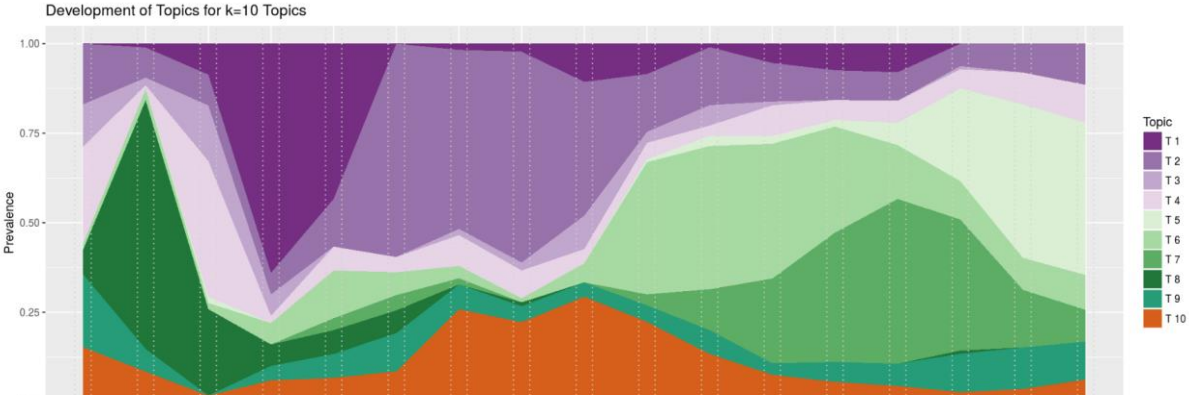

*Figure 4. Discursive landscape: share of UNSC speeches on Afghanistan with dominant topic (T1-T10), 2001-2017. Source: LDA-based topic modeling. Seed year 2017, k=10. Own visualization*

To further understand and be able to assess these topics the LDA produces a list with the words that are most prominent in, and thus define, the topic. Based on the list of the 25 most prominent words in each of these cluster we interpreted and labelled the topics as shown in *Table 2*.

|    | Topic | Characterizing key-words[4] | Peak year(s) |
|----|-------|------------------------------|--------------|
| T1 | Elections | elections, process, electoral, security, election, political, government, Japan, presidential, international. | Main attention 2004-2005; some attention 2009-2010, 2012-2014. |
| T2 | Development | Afghanistan, development, international, efforts, security, community, secretary, assistance, general, afghan. | Main attention 2006-2009. |
| T3 | Ceremonial 1 (Undefined) | will, council, one, time, need, must, many, members, us, even. | Little attention overall. |
| T4 | Drugs | drug, Afghanistan, drugs, trafficking, narcotics, production, counter, opium, cultivation, crime. | Clear peaks 2001 and 2003. |
| T5 | National Peace Process | peace, government, support, process, afghan, Afghanistan, unity, national, conflict, attacks. | Main attention 2015-2017. |
| T6 | Transition | Afghanistan, afghan, will, transition, process, European, union, conference, regional, support. | Main attention 2010-2014. |
| T7 | Women/ Human Rights | Afghanistan, women, rights, security, government, afghan, human, national, progress, 2014. | Main attention 2011-2016. |
| T8 | Reconstruction | reconstruction, police, administration, ISAF, Bonn, Jirga, million, Afghanistan, assistance, security. | Clear peak in 2002, some attention until 2006. |
| T9 | Regional Security | Afghanistan, Pakistan, people, Taliban, peace, terrorism, Iran, India, security, international | Main attention 2001, 2015-2017; some attention overall. |
| T10 | Ceremonial 2 (Broad Issues) | united, nations, afghan, will, must, support, new, work, also, international. | Main attention 2007-2011. |

Table 2.

*Overview of topics and corresponding temporal prevalence. Keywords provided through the LDA algorithm, topic title are authors' summary of the full keyword list per topic, and peak years are based on the data reflected in Figure 4.*

---

[4] For each topic these are the first 10 words of a total of 25 words that the LDA algorithm identifies as defining the topic.

As noted before, besides the temporal anchor, the corpus also provides us with a speaker-affiliation anchor. This means that, more than changes in the relative relevance of topics over time, we can also analyze the speaker network structure underlying the data: we can see which speakers placed their emphasis on similar (or different) topics. In other words, the corpus allows for calculating the aggregated assignment of a speaker (usually representing a country or an UN sub-organization) to a topic reported by the LDA-algorithm across all speeches. An edge between a country or speaker-affiliation is created as soon as one country or organization is assigned to any LDA-topic to some extent. Such a network tends to be near-complete as most countries will have some documents associated with a minimal assignment to nearly every topic. For further analysis, a filtering was therefore applied based on the edge weight measuring the strength of ties between countries and topics. In general, the analysis of the resulting topic-country network can then be conducted on two levels. First, there is the overall network with all countries and all topics. It allows illustrating which topics and which actors are most relevant (in terms of quantity) in the overall discourse. Second, the sub-networks of individual countries can be analyzed in more detail. This allows identifying groups of countries which are interested in similar topics (see *Figure 5*).

*Figure 5* illustrates the overall network structure which shows the topic association of the country-speakers at a 25% analysis level. That means that ties between countries and topics in this network are only kept if countries have an aggregated assignment to a particular topic which is at least 25% of the maximal aggregated assignment of all countries. Country-topic-connections that show less than 25% overall aggregated assignment to a topic are not visualized. Thus, in *Figure 5*, only strong associations between a country and a topic are displayed, which also reduces the number of speaker-affiliations from 103 in the full corpus to the 18 most relevant speakers in the UNSC debates on Afghanistan.

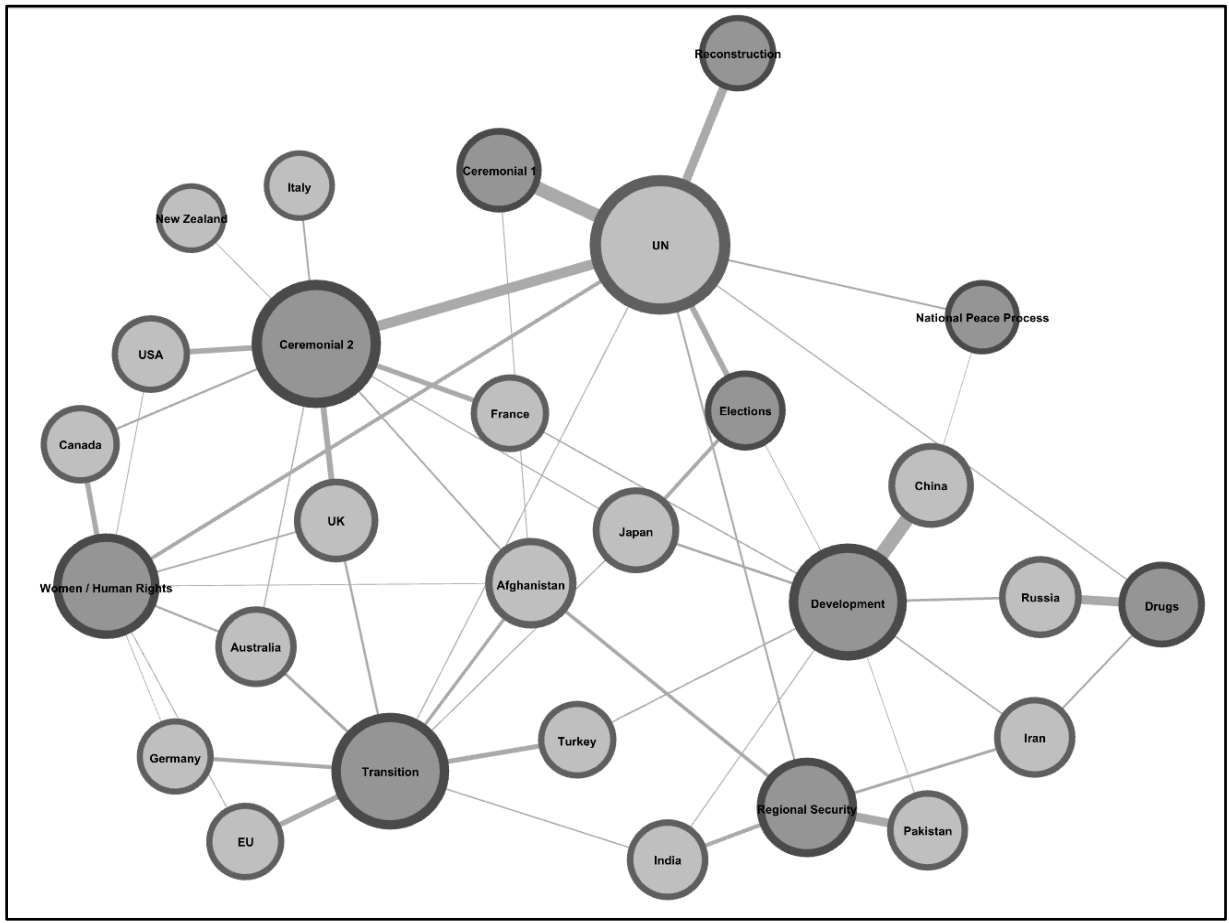

*Figure 5. Two-mode speaker-topic network at 25%-level assignment (=stronger ties), with isolates removed.*

The visualization was produced with the Force2 Layout algorithm in Gephi 0.9.2, with node size representing weighted degree centrality. Nodes in dark grey are topics identified through the LDA; those in light gray are speakers contained in the corpus. Following intuition, edge weights represent the aggregated assignments of all speeches of a country to a topic. The size of a node is determined by a weighted normalized degree centrality, aggregating weights of all edges connected to that node and normalizing the aggregate with respect to aggregates of all other nodes of the same category in the filtered network. That means, for a topic-node, weights of all of its edges are summed up and normalized with respect to all other topic-nodes. Hence, its size represents the relative importance of that topic compared to all topics of the filtered network. For a country-node, its size shows the countries' relative dominance compared to all other country-nodes in the filtered network towards the depicted topics accordingly.

In what follows we use both anchors – yearly topic strength and speaker/country affiliation – to validate the topic assignment produced by the LDA. In a first step, this means we will assess the topics and their temporal changes (as presented in *Figure 4* and *Table 2*) in light of an expert summary of academic reflections of the situation and intervention in Afghanistan: Are major events and slower transitions reflected in the topics, and do we see topics taking dominance in the years that we would expect them to be dominant? In a second step, we assess the overall network structure and zoom in on two particular topics (drugs and women/human rights) to see if the LDA picks up speaker-topic affiliations and whether these, in light of our current understanding of state foci and relations, make sense.

**5. External validation of topics and topic evolution: an expert summary of the conflict**

The conflict in Afghanistan has received quite some scholarly attention over the years. Amongst others, scholars developed models that aim to explain the dynamics of the war (Geller and Alam, 2010), employed case study methodologies to study EU-sponsored police reform in Afghanistan (Eckhard, 2016a, 2016b) and, more recently, conducted quantitative analyses of parliamentary speeches on Afghanistan in Germany and Canada (Lagassé and Mello, 2018). One mode for assessing the validity of the LDA topic categories and their changes over time, is thus to see if they correspond to previous scholarly assessments and descriptions of the conflict (e.g. Berenskoetter and Giegerich (2010), Giustozzi (2016) and Farrell (2018)).

A close reading of this literature allows us to divide the development of the conflict and corresponding intervention into four broad phases: Phase 1 (2001-2003), peace negotiations and reconstruction during the first two years after the intervention; phase 2 (2004 to 2008/09), the consolidation of development assistance and reconstruction; phase 3 (2010 – 2014), the strategic shift towards counterinsurgency, training of Afghan security forces and the subsequent withdrawal of international combat forces; and phase 4 (2015-2017), the invigoration of Taliban resistance with the Afghan state slowly losing control over different provinces. *Table 3* shows that these four phases indeed correlate with the dominant topic(s) for each year as identified by the LDA based topic modelling. However, to provide further context, we summarize the evolution of the Afghan conflict and the activities of the international community in the identified phases and explain how they relate to the different topics and our expectations. In relation to each phase we proceed as follows: we first briefly summarize what defined that phase (i.e. what happened), we then formulate what we would expect to (or should) see in the LDA topics, we then verify whether we can indeed see that the topics reflect the descriptions.

| Year | Rank 1 | Rank 2 | Rank 3 |
|---|---|---|---|
| 2001 | T4 - drugs | T9 - regional security | T2 - development |
| 2002 | T8 - reconstruction | | |
| 2003 | T4 - drugs | T8 - reconstruction | |
| 2004 | T1 - elections | | |
| 2005 | T1 - elections | T2 - development | T6 - transition |
| 2006 | T2 - development | | |
| 2007 | T2 - development | | |
| 2008 | T2 - development | | |
| 2009 | T2 - development | T10 - ceremonial/ broad | |
| 2010 | T6 - transition | T10 - ceremonial/ broad | |
| 2011 | T6 - transition | T2 - development | |
| 2012 | T6 - transition | T7 - women/human rights | |
| 2013 | T7 - women/human rights | T6 - transition | |
| 2014 | T7 - women/human rights | T6 - transition | |
| 2015 | T7 - women/human rights | T5 - national peace process | |
| 2016 | T5 - national peace process | T7 - women/human rights | |
| 2017 | T5 - national peace process | T2 - development | |

*Table 3. Topic prevalence for each year (ranking of topic(s) accounting for a total of at least 50% of attention. Color represent phases.).*

*Peace negotiations and reconstruction (Phase 2001-2003):* Before 2001, the Security Council was mainly concerned with Afghanistan from the perspective of regional security. This, because of the perceived threat of the Taliban regime under Mullah Omar, who offered sanctuaries for terrorist groups and was involved in the cultivation of poppy (Felbab-Brown, 2006). After 2001, when the US-led coalition overthrew the Taliban regime, a coalition of states and international organizations tried to guide the country towards democracy, peace and security. Different from Kosovo and East Timor (both since 1999), the international community decided not to mandate a major UN peacekeeping force with the reconstruction of Afghanistan but to initiate a political process. This process was to result in a local power sharing deal. What has become known known as the 'Petersberg process', and the thereto related 'Bonn agreement', led to a so-called 'Loya Jirga' – a great assembly – that installed Hamid Karzai as transitional president until elections could be held in 2004. Moreover, during these first years the international community convened a reconstruction and aid conference in Tokyo, January 21-22, 2002. As outlined by the 'lead nation approach', individual donors assumed responsibilities for rebuilding various sectors of the Afghan state (policy, judiciary, bureaucracy). As such, considering what we know from this first phase, we expect the initial focus of discussions in the UNSC to be on topics related to the Taliban threat, the search for a political solution to the conflict and the development of a state-building and reconstruction strategy.

The automated LDA identifies the following topics as most prominently discussed in 2001–2003. For 2001 these are: Topic 4 (drugs) with 27%; topic 9 (regional security) with 20%; and topic 2 (development) with 17%. In 2002 as single topic was dominant, topic 8 (reconstruction, police and administration) with 69%. In 2003, the focus was again on topic 4 (drugs, 38%) and 8 (reconstruction, 24%). The prevalence of topics 4 and 9 in 2001 reflects the pre-9/11 debate that still dominated most of 2001 and that was mainly concerned with the security threat. Drug-trafficking (topic 4) was the main source of income for the Taliban regime and regional security (topic 9) reflects the security threat of the regime and local terrorist groups. The dramatic shift of attention to topic 8 in 2002 and 2003 reflects the search for a reconstruction strategy for Afghanistan after the US-led coalition had overthrown the Taliban regime (without involving the UNSC): Speech acts belonging to topic 8 for instance often mention the 'Bonn-agreement' and the 'loya jirga'. The latter was to be held in June 2002 and led to establishment of the transitory government and the first elections in 2004 and 2005. As an example of how this is empirically reflected in the topic, speakers would for instance stress the need to build political institutions, 'to [re-build] the economy and society of that war-ravaged land' and to support the 'international community in helping to rebuild Afghanistan's army and police' (Mr. Khalid, Pakistan in UNSC, S/PV.4541, 2002). In all we see that the outcome topics of the LDA indeed reflect what we may expect in the first phase of the intervention.

*Consolidation of development assistance (Phase 2: 2004-08)*: The first years of this phase were a temporary phase of peace and low levels of violence, despite the fact that US-American forces were engaging Al Qaeda and Taliban fighters as aggressors. However, the initial unwillingness of the international community to deploy their International Security Assistance Force (ISAF) outside Kabul, coupled with the initial weakness of the new Afghan state, created a security vacuum throughout the country that Taliban groups used to restructure. When ISAF finally expanded to the calmer North (2003) and specifically to the more contested South (2006), they met fierce resistance by local insurgents linked to the Taliban. At the same time, following the political process instigated by the Loya Jirga, Hamid Karzai was elected as first president of Afghanistan in 2004 after regular presidential elections, and parliamentary elections followed in September 2005. Moreover, and despite an increase in security incidents in the country – including an attempted assassination of President Karzai in Kabul in April 2008 – 'development' remained the dominant concern: In 2006 a donor conference was organized in London, the 'Afghan Compact', and a later analysis by the International Crisis Group (2008, p. 7) recalled:

> "At the time the Compact was presented, the seriousness of the insurgency had still not been widely recognised, and commitments focused on moving from stabilisation to state building and reconstruction; there was even talk of the U.S. withdrawing some troops."

Considering the presidential and parliamentary elections in 2004 and 2005 respectively, and the described relative neglect of the Taliban insurgency, we expect the LDA of the 2004-2008 period to reflect the elections as well as the further focus on development and reconstruction of an Afghan state (Rubin & Hamidzada, 2007).

The topics identified as relevant by the automated topic analysis coincide with these expectations. Topic 1, 'elections' (see also *Figure 4*), dominated the first two years – with 64% in 2004 and 43% in 2005 – after which the LDA identifies a shift in focus towards topic 2, 'development', from 2005 to 2009 (with 13%, 60%, 50%, 59% and 37% respectively). Speakers within this topic referred more frequently to the 'progress towards the political and economic stabilization' of the country (Mr. Al-Nassar, Quatar in UNSC, S/PV.5385 2006) but also argued that, in light of continuing challenges, 'it is important for the Afghan authorities and the international community to maintain their efforts' (Mr. Lacroix, France in UNSC, S/PV.5581 2006). The dominance of the development topic slowly phases out after 2008, giving rise to two new topics that dominate the subsequent phase.

*Counterinsurgency and exit strategy (Phase 3: 2009-14)*: Although the US military had begun deploying additional troops to ISAF in 2007 (from 40,000 to 50,000), a real strategic shift only occurred after the election of Barack Obama as US president in 2008. He convinced NATO member states to send additional troops which lead to a maximum of 130,000 coalition troops in Afghanistan in the peak year of 2011. Coinciding with the troop surge, the number of coalition force fatalities increased significantly reaching around 500, but up to 700, a year in the period 2009-2011. This led to increased discussions about possible exit strategies. An important element of this exit strategy was to train the Afghan security forces, which also led to a further influx of private military and security contractors on the side of the US, which had more than one hundred-thousand contractors working in different positions as per October 2011 (CENTCOM 2011, p.1).[5] After 2012, and supported by initial military success, the number of ISAF forces was reduced and the mission closed in 2014. Therewith 2014 marks the year in which the responsibility for public security transitioned from the international coalition to Afghan security forces. By that time the force of the Afghan security institutions had grown to around 350,000. Only a small NATO training mission remained to support the Afghan government. Considering the above, 2009 should be about the time for the dominant narrative to shift from development to security. Moreover, we would expect the LDA to pick up on the force transition towards the end of this phase as dominant focus.

Interestingly, and somewhat surprisingly, we do see our expectations largely confirmed – we indeed see that 'transition' becomes visible as a separate topic around this time (topic 6, transitions, is prominent from 2010 to 2014 taking respectively 37%, 40%, 38%, 30% and 15% of the focus) – but the troop surge, counterinsurgency and thereto related security issues are not identified as a separate topic. Instead, 'security' is woven into both the discourse on the transition and the second topic that became prominent around this time: topic 7 on 'women and human rights' (between 2012-2016 with 24%, 36%, 46%, 37%, 16% respectively). Speech contributions within topic 6, 'transition', refer to the slow handing over of domestic security responsibilities to the newly trained Afghan forces and police. An example for a related speech act is the Turkish representative, Mr. Apakan, who said in 2012 "I would like to express our satisfaction with the Afghan-led security sector transition, whereby the Afghan National Security Forces now have full responsibility for nearly half of the population" (Mr. Apakan, Turkey in UNSC, S/PV.6735, 2012). The relevance of topic 7, 'women and human rights', was not foreseen by us but may still be explain in light of the increased violence and troop surge of the time. The growing number of military operations and insurgencies in the country which often, and only partially unintendedly, targeted civilians in general and women in particular.

---

[5] Of these approximately 23,000 were US-Americans, over 50,000 were Afgan-nationals hired by US-American military contractors and another 28.000 were so called 'third country nationals' contracted largely from countries from the Global South (CENTCOM 2011, p.1; see also van Meegdenburg 2017).

For instance, the Canadian Ambassador said in 2013: "The past few months have also seen a marked increase in the number of attacks on civilians, including humanitarian workers, Government contractors and politicians, particularly women" (Mr. Rishchynski, Canada in UNSC, S/PV.7035, 2013). In part because of the relatively surprising appearance of this topic in these particular years we will further highlight the dynamics behind topic 7, 'women and human rights', in relation to its topic-speaker structure later in this analysis.

*Towards a political settlement (Phase 4: 2015-17)*: In the wake of the ISAF withdrawals, and its supersession by the small NATO training mission 'Resolute Support', Taliban fighters intensified their activities and presence throughout Afghanistan. Government forces and the police in many provinces lost control of public security and the country overall became more insecure. Besides increased insecurity this also led to the recognition that a political settlement with the Taliban was necessary for a stable future. Western and Afghan diplomats had already held secret negotiations with Taliban representatives in 2011, brokered by Pakistan and Qatar, but officially the Taliban had been classified as a terrorist organization. Since 2015/16, Afghan politicians openly advertised a power settlement. Up to today, however, such an agreement has not been reached and it is now the Taliban who seek military supremacy. In this final phase we would thus also expect the discourse to be dominated by the described deterioration of security as one topic. The second topic we expect the LDA to have identified is the new political settlement and negotiations with the Taliban.

In relation to the first, we see a continuation of the subsumption of security or, better, the increased insecurity under topic 7, 'woman and human rights'. In relation to the second aspect we indeed see that the automated analysis reflects the developments with the emergence of a new dominant topic after 2015: Topic 5, 'national peace process' (2015-2017: 26%, 42%, 43%). Speech acts within this topic refer to new initiatives linked to the peace process, such as Ambassador Rosselli (Uruguay) who said in 2016 that "Uruguay welcomes the initiatives undertaken to revive the peace process, as well as the agreement reached by the Quadrilateral Coordination Group on Afghan Peace and Reconciliation for the peace talks held in December 2015" (Mr. Rosselli, Uruguay in UNSC, S/PV.7645, 2016). Topic 5 therewith takes over from topic 7, 'women and human rights', despite a still troublesome security situation with daily security incidents and many casualties throughout Afghanistan. Moreover, 2017 saw an increase of topic 2 on development (12%), but also two topics from the pre-09/11 time, drugs (topic 4, 11%) and regional security (topic 9, 11%) which fits in with the renewed focus on a political settlement.

Overall, we find the fit between the topics identified by the LDA and the overall, qualitative and chronological descriptions of the conflict remarkably strong. In line with our first hypothesis (H1), the LDA is able to capture the overall trends and dynamics of the intervention from the (written out) contributions to the UNSC: The topics identified by the LDA follow logical shifts in dominance and reflect key events and periods. At the same time, the analysis also identified topics that we had not initially thought off. Especially topic 7, 'woman and human rights', was not predicted by us on the basis of the qualitative descriptions of the conflict and intervention. This illustrates that automated topic modelling of a text body can be an analytic end in itself leading, on the one hand, to a valid depiction of a discursive landscape and, on the other hand, to the possible identification of topics one might not have foreseen. Yet, it also illustrates that automated analysis of a vast text corpus can aid other studies by, for instance, identifying interesting and surprising topics (discourses) for further research.

That said, it should be noted that although the outcome topics of the LDA are interpretable for people with knowledge of the situation and the chronology of the intervention, non-experts (lay-people) might have found it challenging to label and validate the topics based on 25 keywords and the evolution of topics over time. This is an issue that may become even more challenging, or require even larger collaborative efforts, when the corpus-based research is extended to larger and multi-issue corpora of diplomatic speech.

**6. External validation by speaker-topic network: a text-as-network approach**

Focusing on the second anchor – speaker affiliation – provides us with a second means of validation. To see whether the LDA topic model gives us meaningful topic categories, and thus a meaningful summary and clustering of the speeches, we analyze whether we can differentiate speakers (country and UN representatives) on the basis of the topics they predominantly address. Here we would first of all expect to find that different countries place different emphasis. This means that we should see edges of different strength (visually: thicker) between certain countries and certain topics. Secondly, we would expect that certain groups of states – states with strong interstate relations such as the EU or NATO member states, or certain classical divides as between the P3 (USA, UK, France) and Russia and China – to be visible in a network analysis based on the LDA. In what follows we provide a network analysis of the LDA topic categories and speaker affiliations to test these expectations. We first zoom in – at 15% speaker-topic assignment level with more countries included than in *Figure 5* – on two particular topics and the speakers affiliated to these topics. Emphasizing the different country foci, we selected two topics that were placed at different ends of the two-mode network, connected only through ties with the UN node in *Figure 5*: topic 4 "drugs" and topic 7 "women and human rights". These topics should therefore represent geopolitically differences in attention, while they also point to very distinct and most discourses with little overlap. Thereafter, we zoom out back to the 25% speaker-topic assignment, but in a one-mode country-country projection, to study how countries cluster according to strong co-affiliation to common topics they speak about.

*Country-topic priorities topic 4 – 'drugs'*: This topic addresses the poppy cultivation in Afghanistan, drug trafficking by, and as a source of income for, the Taliban and Al-Qaida and other drug related issues. As can be seen in *Figure 6,* the topic is dominated by Russia with secondary ties to the UN and Iran. What is not visible in this network (at a 15% analysis level), but what would be visible if we lower the bar is that, beyond Russia, other members of the Russian led Collective Security Treaty Organization (CSTO) were also active speakers on this topic when they had a seat at the UNSC as well as other regional actors such as Pakistan and India.[6] With the exception of the UN and France, this means 'drugs' is predominantly a Russian topic with regional secondary speakers. How can we understand this?

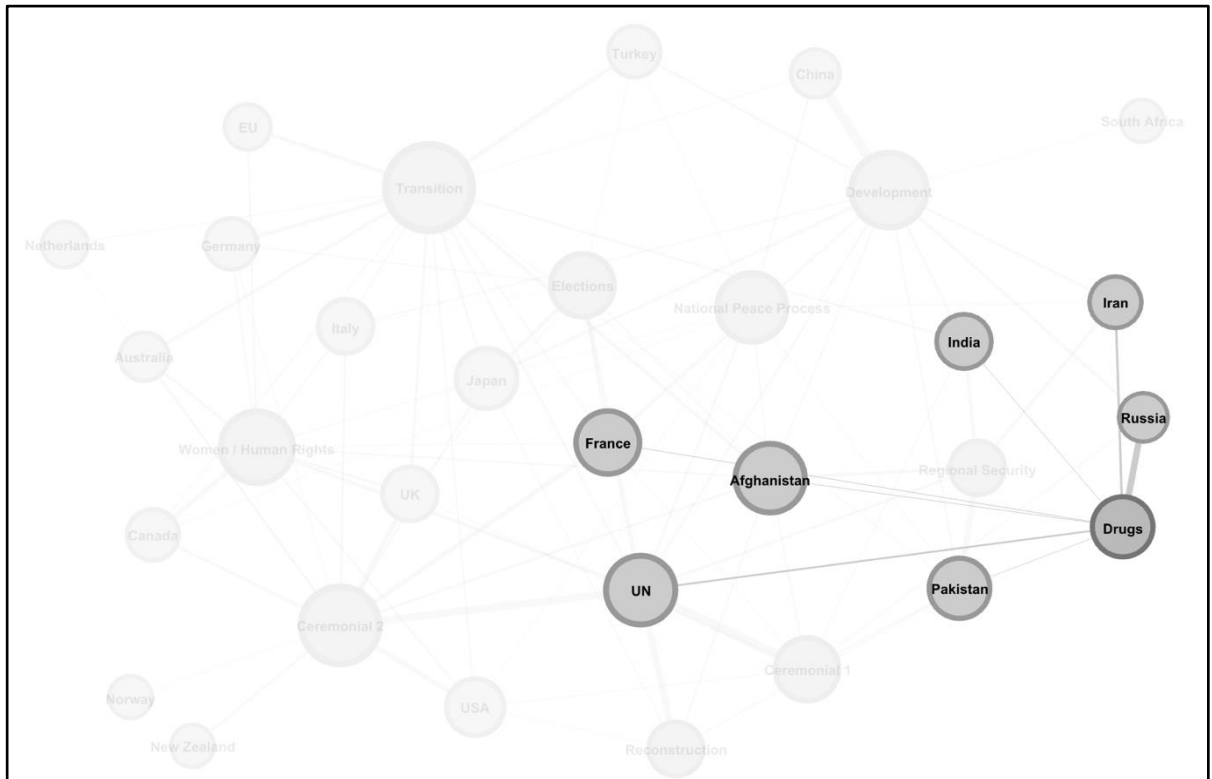

*Figure 6. Speaker-topic network at 15% assignment, focus on topic 4: "drugs", isolates removed.*

For these states, narcotics from Afghanistan constitutes a domestic security threat both because of the large supply of opium and because the illicit trafficking fuels a major organized crime industry. Initially, in the early days of the intervention (2001-2003, see *Figure 4*), drugs were a major topic. At that point in time the topic was also addressed by many Western states resulting in, and from, a proposal to try and eradicate the production of opium in Afghanistan as an important source of income for the Taliban and Al Qaeda. However, it is interesting that, whilst opium production kept and keeps increasing, this topic afterwards lost priority. Whilst in 2003 most speakers emphasized, as did for instance Mr. Vohidov from Uzbekistan, that "recent events have provided irrefutable proof of the link between international terrorism and the drug trade that fuels it" (Mr. Vohidov, Uzbekistan in UNSC, S/PV.4774 Resumption1, 2003) the prominence of this topic dwindles quickly. As John Walters and David Murray concluded recently in an article in Foreign Policy: "Since the liberation of Afghanistan, U.S. and allied policy has failed to press the elimination of opium production as a necessary priority" (Walters and Murray, 2017) and opium production in the country was at an all-time high in 2017 (UNODC, 2017). This lack of attention is visible in *Figure 4* but becomes even clearer when looking at the network structure of this topic in *Figure 6*. This structure makes clear that besides the UN, which in this case often is represented by the executive director of the United Nationals Office on Drugs and Crime (UNODC), Yury Fedotov, it is primarily Russia who keeps the topic of drugs on the agenda. In 2010, for instance, with Topic 4 at its lowest relative weight, Russia was the only country which prominently addressed drugs and opium in its contributions.[7] This illustrates that the country was relatively isolated with its concerns and in emphasizing drugs as a security issue.

---

[6] The other members of the CSTO are: Armenia, Azerbaijan, Belarus, Georgia, Kazakhstan, Kyrgyzstan, Tajikistan, and Uzbekistan.

In short, zooming in on topic 4 provides us a with a first clear indication that the network analysis validates the LDA as the network-structure that can be visualized on its basis displays a logical speaker-topic connection in relation to this topic.

*Country-topic priorities topic 7 – 'women and human rights'*: The topic addresses, broadly, the issue of human rights and the position of women in the Afghan society at a time when fighting and civilian casualties in Afghanistan intensified. The focus on women as part of this topic probably relates to UNSC Resolution 1325 on women, peace and security which was adopted on October 31, 2000 and was followed by range of follow-up resolutions in particular between 2008-2015 (Thomson 2018, 3). In short, Resolution 1325 refers to an 'urgent need to mainstream a gender perspective into peacekeeping operations' and stresses 'the importance of [women's] equal participation and full involvement in all efforts for the maintenance and promotion of peace and security'. Two observations stand out when assessing the sub-network on topic 7: First, the central role of the UN and, second, that the question of women is addressed almost exclusively by Western states with Canada in the lead (see *Figure 7*).

---

[7] In that year only eight speeches prominently addressed the issue of drugs and drug trafficking in Afghanistan (with a topic score of more than .20) and all where from the Russian delegation.

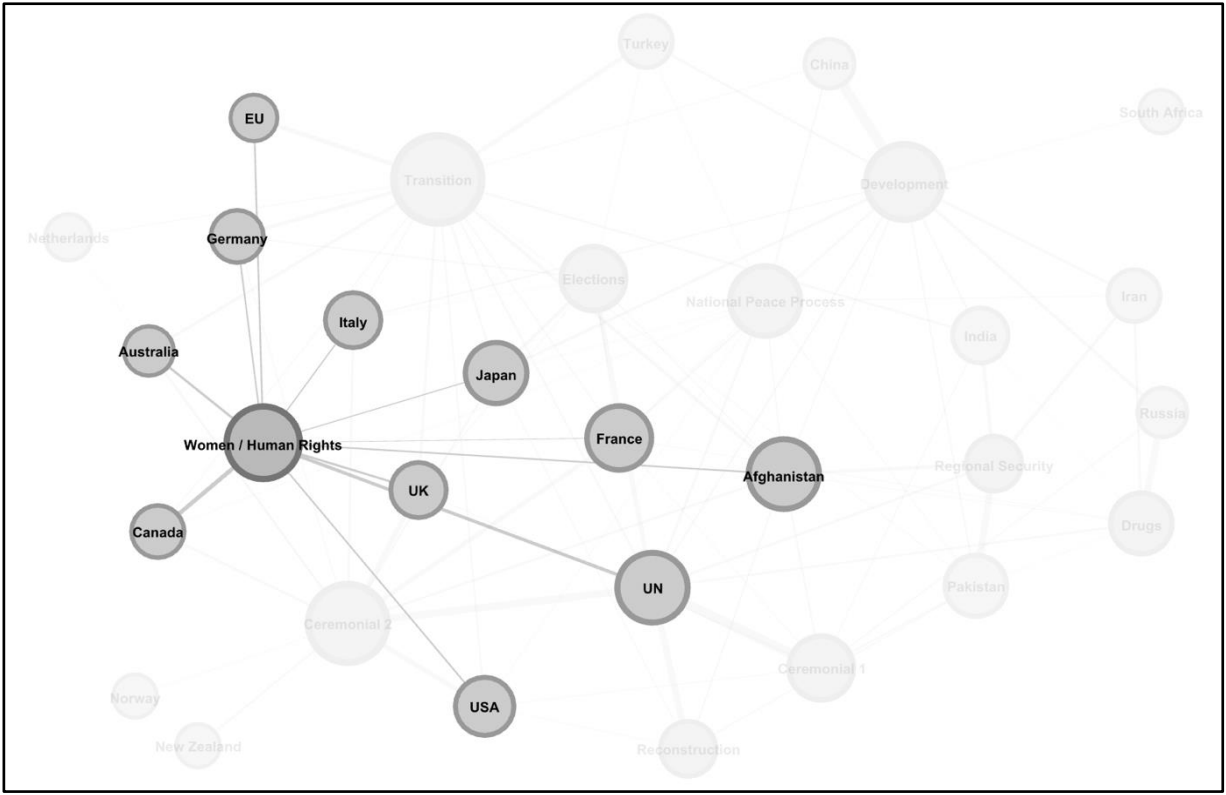

*Figure 7. Speaker-topic network at 15%-level assignment, focus on topic 7 "women and human rights", isolates removed.*

Firstly, from the temporal analysis of the LDA topic categories we know that the topic 7 on women and human rights became relevant in the UNSC discourse on Afghanistan by 2012-13 only (see *Figure 4*). Whilst the original resolution on "gender mainstreaming" was passed in 2000, the first contribution that truly addressed the topic (above a relevance threshold of .20) was in 2005 by Canada. In 2006 three interventions surpassed a .20 relevance threshold. These interventions were by Canada, Denmark and Iceland. That member states are increasingly referring to women as a topic in the Afghanistan debate follows a pattern also visible in the annual UNSC Resolutions that extend UNAMA. Whereas the original resolution that established the mission did not mention women at all (Res. 1401, 28.03.2002), the number of times the term appeared in the resolutions increased steadily since then – Res. 1589 in 2005: 1; Res. 1806 in 2008: 5; Res. 1917 in 2010: 10; Res. 2041 in 2012: 21; Res. 2274 in 2016: 46. This reflects a steady process of gender mainstreaming in the documents of the UNSC as intended by the original by Resolution 1325 and following the Canadian lead on addressing the position of women in the UNSC.

Secondly, it is particularly notable that topic 7 is almost exclusive addressed by Western countries, most specifically Canada with the highest edge weight in its tie to the topic compared to all other speakers referring to this issue. Canada is the lead nation of the *ad hoc* and informal group of states known as the 'Friends of [Resolution] 1325'.[8] At the same time, Canada is the only country that was *not* a formal member of the UN Security Council during the time analyzed in this paper. Instead, Canada participated under Article 31 of the UN Charter that allows any member of the UN to participate, without the right to vote, in the discussion before the Council. Only few countries make use of this right and it is interesting to see that Canada actively shaped the international discourse on Afghanistan by putting women's rights on the agenda. One example for a speech that prioritizes women's rights is the 2017 statement by Canadian Chargé d'Affaires Mr. Bonser: "There are three themes that I will address today. First, the full and equal participation of women in all facets of Afghan society is essential. (…)" (Mr. Bonser, Canada in UNSC, S/PV.8147). Although a more nuanced analysis would be necessary to better understand the role and possible success of Canada as norm-entrepreneur and agenda setter of women's rights, the fact that topic 7 is an almost exclusively 'Western' topic is in line with expectations and further validates the LDA on which the network analysis is based.

*Overall country clusters* – Lastly, besides zooming in on particular topics the network representation of the discursive landscape allows us to identify particular clusters of countries that are most strongly connected by the topics they jointly address. To identify these clusters, we conduct a one-mode country-country projection of the two-mode network presented above. For this we used the "MultiMode Networks Projection" plugin for Gephi (assigning a type "country" or "topic" to each node). To identify clusters or groups of states that are most strongly connected by the topics in the networks, we apply the community detection algorithm ('modularity') implemented in Gephi (Blondel et al. 2008; Lambiotte et al. 2009) at two different sensitivities (1.0 and 0.33; randomization and edge weight applied), with lower values increasing the number of communities detected. The resulting two one-mode networks where node color represents affiliation to the same automatically detected group underlines that, starting from an unsupervised LDA-based topic assignment and a semi-automatic speaker-to-speech identification, we find traditional geopolitical divisions and regional alliances with an automatic community detection algorithm. *Figure 8* shows the one-mode country-country network with a low number of communities (two communities), showing a major East-West divide beyond just a simple Russia versus the rest in the UNSC, basically clustering Russia and China as Permanent Five members with the closest distance to Afghanistan alongside countries with regional security interests related to Afghanistan (Iran, Pakistan, India, and Turkey).

---

[8] Other members are: Australia, Austria, Belgium, Brazil, Canada, France, Germany, Italy, Japan, Liechtenstein, Norway, Portugal, the United Kingdom, Uruguay and Switzerland.

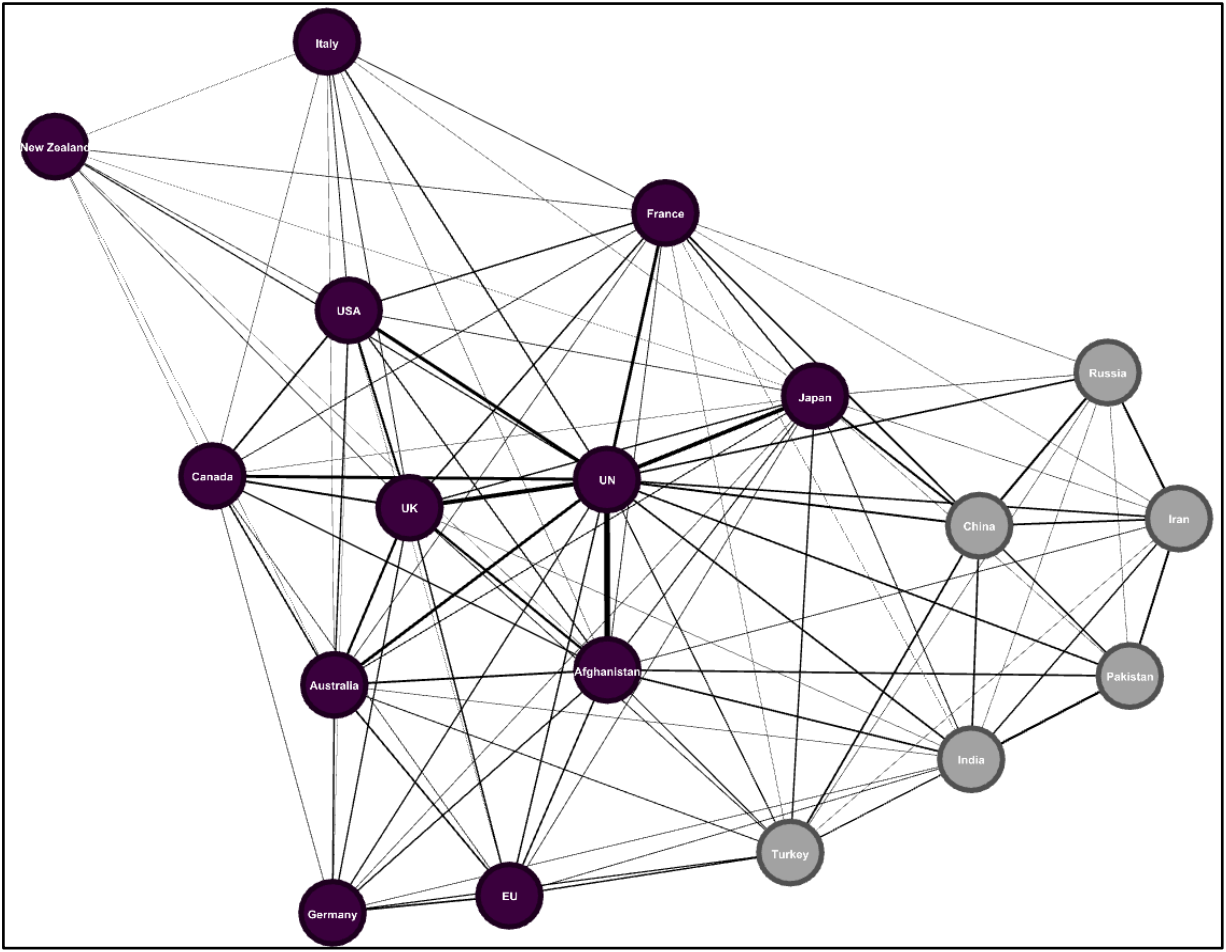

*Figure 8. One-mode country-country network of UNSC debates on Afghanistan with low number of communities.*

*Figure 9* with a higher number of communities provides a more nuanced picture of topic-related ties between different countries. At the center of the network one finds the UN and Afghanistan, because speakers representing these actors have been talking about the greatest variety of topics, thereby gaining a central position and linking the other countries with more limited interests through their broad interest in most topics. At the same time, one finds that Western actors like Germany, the EU and Australia seem to be focused on different topics than other Western actors like the USA, UK or Canada. At the same time, states that are known to have quite different interests such as India and Pakistan are shown to talk about similar topics. The same is true for the cluster formed by China, Russia and Iran.

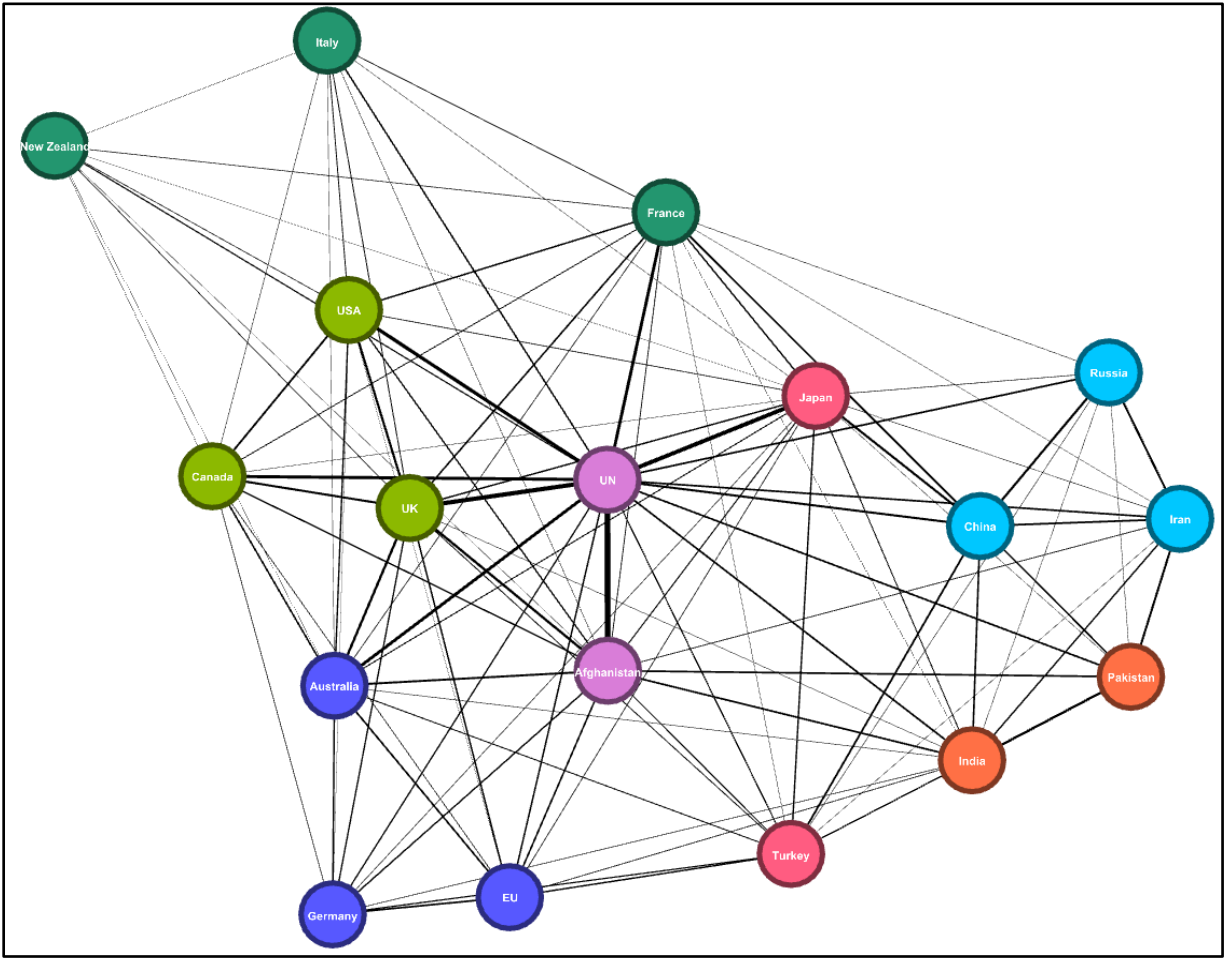

*Figure 9. One-mode country-country network of UNSC debates on Afghanistan with high number of communities.*

The above discussion and network visualizations show that, beyond a qualitative validation of the topics and the changes in their prominence over time, a network analysis based on the LDA topics provides a second, strong means of validating the automatically produced topic-categories. The country foci – Russia and regional actors on 'drugs' and Canada and other Western speakers on 'women and human rights' – as well as the overall speaker networks presented in *Figures 8 and 9* represent traditional or otherwise comprehensible divides in the UNSC and beyond. It also allows us to clearly illustrate that different states do place different foci in their UNSC contributions and that these differences are picked up by the LDA algorithms. This means that, in all, we have strong reasons to believe that the topics produced by the automated LDA represent valid aspects of the debates on Afghanistan.

## 7. Conclusion

When states intervene in the UNSC they are part of, and (re-)create, a (geo)political discursive landscape. This landscape is shaped by the articulations of states' particular interests and concerns, often based in and reproducing structural divides between states in the international system, and thus in the UNSC. In this paper we presented and validated LDA based topic models to study and visualize this landscape in relation to the conflict and intervention in Afghanistan. A method, we would like to stress here, we see as complementary, not challenging, qualitative approaches to discourse.

As we argued earlier, two of the major advantages of this approach to studying text and discourse is that it enables analysis beyond the point where qualitative approaches reach their limits, and that it allows for different foci and forms of visualization and interpretation. Firstly, by means of automated, quantitative text-as-data approaches we are able to process and analyze large corpora of text. These corpora may span multiple years (or even decennia), include numerous and diverse actors and may be composed from multiple international fora or individual speeches and written contributions. Especially when pre-processing is aided by clear document structures (such as is the case at the UNSC) there are virtually no limits. This does not allow for the type of fine-grained analysis that is generally accomplished by qualitative, in-depth analysis of actor particular discourses, but that is not the aim. Instead, this type of quantitative analysis allows us to lay bare the overall structure of an international discourse. And so, secondly, quantitative approaches allow us to present and understand discourses in different ways. Rather than focusing on the construction and (re-)production of particular meanings – however important – it can help us to visualize and present the relative relevance of a particular topic, the changes therein over time or between fora and the underlying network-structures of the *overall* discourse.

Moreover, an important element of the complementarity of quantitative approaches to discourse is that they can facilitate case selection for further, qualitative analysis. As we showed in relation to topics 4 ('drugs') and 7 ('women and human rights') quantitative approaches are able to identify dominant speakers as well as the time-spans during which a particular topic was salient. While the default option may be to study the 'great powers', especially the US considering it "looms large over the Security Council" (Puchala 2005, 574), the discursive landscape projected here is much more complex, identifying different speakers as dominant or important. Based on the topics and their perhaps unexpected salience at a particular point in time, or based on interesting and surprising speaker affiliations to particular topics, further qualitative analysis may reveal elements that remain concealed by quantitative approaches.

This last point also directs attention towards a particular weakness of quantitative approaches for unsupervised topic assignment: As presented here, LDA topics do not reveal *how* a particular country spoke about a topic, nor *why* it did so. It only tells us *that* they did. Why did Russia dominate the drug topics and why is Canada in the lead when it comes to women and human rights? And, equally important, did Canada refer to women and human rights positively (as something worth supporting) or negatively (as something to be ignored)? Intuition tells us the former is true but strictly speaking the subsumption of Canadian contributions in the topic and the prominent place it takes in the network does not give us any indication that this so. In fact, to verify we would need to look into the actual speeches; we would need to supplement the analysis with a qualitative evaluation of the discourse itself or additional quantitative methods such as sentiment analyses. In the online tool we provide to study the dataset, the former is easily possible.[9] Looking into a particular topic one is presented with all the contributions that constitute the topic, ordered from more to less relevant. Clicking on any of these contributions opens the respective contribution after which it can be read or searched using key-words.

In all, in this paper we showed how quantitative text analysis – through non-supervised LDA-based topic modelling – can represent the structure of a discourse created over the course of almost two decades. We were able to show how such topic modelling allows tracing the shifts in dominant topics over time, providing tools to identify changes in the landscape. Moreover, we demonstrated how such a large text corpus can be used to analyze the overall, international discourse of a key political issue – in our case the conflict and intervention in Afghanistan – and that this allows us to identify and subsequently focus on aspects of the discourse that 'stand out' in one way or another. Lastly, we argued and illustrated how such a quantitative approach, combining LDA-based topic modeling and speaker-topic structures, can complement qualitative analysis. And, in relation to this last point, we would like to explicitly invite qualitatively oriented scholars to use (or challenge) our analysis and to assess the data we provide from different angles.

---

[9] See ... (censured for peer-review purposes).